\documentclass[aps,prd,twocolumn,showpacs,floatfix,superscriptaddress,
preprintnumbers,nofootinbib]{revtex4}
\usepackage{graphicx,amssymb,url}

\newcommand{\be}{\begin{equation}}
\newcommand{\ee}{\end{equation}}
\newcommand{\ba}{\begin{eqnarray}}
\newcommand{\ea}{\end{eqnarray}}

\def\lsim{\raise0.3ex\hbox{$\;<$\kern-0.75em\raise-1.1ex\hbox{$\sim\;$}}}
\def\gsim{\raise0.3ex\hbox{$\;>$\kern-0.75em\raise-1.1ex\hbox{$\sim\;$}}}

\begin{document}

\title{Superdense cosmological dark matter clumps}

\author{V.~Berezinsky}
 \affiliation{INFN, Laboratori Nazionali del Gran Sasso, I--67010
  Assergi (AQ), Italy}
 \affiliation{Center for Astroparticle Physics at LNGS (CFA), I--67010
  Assergi (AQ), Italy}
 \affiliation{Institute for Nuclear Research of the Russian Academy of
 Sciences, Moscow, Russia}
\author{V.~Dokuchaev}
 \author{Yu.~Eroshenko}
 \affiliation{Center for Astroparticle Physics at LNGS (CFA), I--67010
  Assergi (AQ), Italy}
 \affiliation{Institute for Nuclear Research of the Russian Academy of
 Sciences, Moscow, Russia}
\author{M.~Kachelrie\ss}
 \affiliation{Institutt for fysikk, NTNU Trondheim, N--7491 Trondheim,
  Norway}
\author{M.~Aa.~Solberg}
 \affiliation{Institutt for fysikk, NTNU Trondheim, N--7491 Trondheim,
  Norway}

\date{February 18, 2010}

\begin{abstract}
The formation and evolution of superdense clumps (or subhalos)
is studied. Such clumps of dark matter (DM) can be  produced by many 
mechanisms, most notably by spiky features in the spectrum of 
inflationary  perturbations and by cosmological phase transitions. 
Being produced very early during the radiation dominated epoch,
superdense clumps evolve as isolated objects. They do not belong 
to hierarchical structures for a long time after production, and 
therefore they are not destroyed by tidal interactions during the formation
of larger structures. For DM particles with masses close to the
electroweak (EW) mass scale, superdense clumps evolve towards a power-law 
density profile $\rho(r) \propto r^{-1.8}$ with a central core. 
Superdense clumps cannot be composed of standard neutralinos, since
their annihilations would overproduce the diffuse gamma radiation.
If the clumps are constituted of superheavy DM  
particles and develop a sufficiently large central density, the evolution 
of their central part can lead to a 'gravithermal catastrophe.' In such a
case, the initial density profile turns into an isothermal profile
with $\rho \propto r^{-2}$ and a new, much smaller core in the center. 
Superdense clumps can be 
observed by gamma radiation from DM annihilations and by gravitational 
wave detectors, while the production of primordial black holes and cascade 
nucleosynthesis constrain this  scenario.
\end{abstract}


\maketitle

%%%%%%%%%%%%%%%%%%%%%%%%%%%%%%%%%%%%%%%%%%%%%%%%%%%%%%%%%%%%%%%%%%%%%%%%%%
\section{Introduction}
\label{intro}

Gravitationally bound structures in the universe have developed from
primordial density fluctuations $\delta(\vec x,t)=\delta\rho/\rho$
that in turn  were produced at inflation from quantum fluctuations. 
In the standard approach to inflation, the spectrum of these primordial
fluctuations   
has a nearly scale-invariant form, $P(k)\equiv\delta^2_k\propto
k^{n_p}$ with $n_p\simeq1$.  During the radiation-dominated (RD) epoch
fluctuations grow slowly, $\delta_k\propto\ln(t/t_i)$, while they grow as 
$\delta_k\propto (t/t_{\rm eq})^{2/3}$ after the transition to the 
matter-dominated (MD) epoch  at $t=t_{\rm eq}$. 
Gravitationally bound objects are formed and detach from the cosmological 
expansion, when fluctuations enter the 
non-linear regime $\delta\geq1$.  The non-linear stage of fluctuation growth 
has been studied both analytically~\cite{Gunn77,Ber85,ufn1,SikTkaWan96} and in
numerical simulations~\cite{Gott75,NFW,makino,moore,JingSuto} for the formation of galaxies
and structures on larger scales.  The density profile in the inner part of 
dark matter (DM) halos is given by
$\rho(r) \propto r^{-\beta}$, with $\beta \approx 1.7 - 1.9$ in analytic
calculations \cite{ufn1}, $\beta =1$ in the simulations of Navarro, Frank and 
White \cite{NFW} and $\beta =1.5$ in the simulations of Moore 
{\it et al.\/}~\cite{moore} and Jing and Suto~\cite{JingSuto}.  

The smallest DM objects in the universe, which we shall call 
clumps or subhalos, are produced first. The evolution of DM clumps has been
studied in Ref.~\cite{bde03} in the hierarchical model in which due to the 
merging of objects a small clump is hosted by a bigger one, the latter is submerged into
an even bigger one, etc. The important observation of \cite{bde03} was the
role of tidal interactions, which fully disrupt most clumps. 
The survived clumps can be further destroyed in the Galaxy by tidal
interactions in the Galactic plane, near the Galactic center, and in
collisions with stars in the halo (see \cite{BDE08} for a review). The
characteristic feature of these processes of disruption is that the core
of a clump survives and thus the gamma signal from DM annihilations in clumps changes 
only mildly~\cite{BDE08}. A statistical approach to the search for
galactic small-scale substructures has been recently proposed 
in \cite{KamKou08,KamKouKuh10}. 

The mass spectrum of DM clumps has a low-mass cutoff $M_{\min}$ due to
the leakage of particles from a clump. This mass is strongly 
model dependent: It depends on the leakage mechanism (free streaming,
collisional damping, etc.), on the properties of the DM particles and 
the resulting decoupling temperature and others. Therefore, the 
predicted $M_{\min}$
varies for neutralinos in the minimal supersymmetric standard model 
from $10^{-7}$ to $10^{-5} M_{\odot}$~\cite{bde06,cutweak}.

We have described above the standard cosmological scenario for the clumps. 
In non-standard scenarios the properties of  DM clumps can be very 
different. In Ref.~\cite{kt}, {\em isothermal perturbations\/} 
in the DM density were considered within the framework of a
spherical collapse model. Perturbations collapse in the
RD epoch and produce superdense DM objects.  
Another possibility for the production of superdense clumps is given by 
a spiky spectrum of  perturbations~\cite{star,GelGon08,Scott}.
The general idea common to these scenarios is that there exists a 
spike on top of a scale-invariant power-law spectrum of  perturbations
which results in the production of dense clumps in a very early cosmological
epoch. In this work, we consider in contrast to~\cite{kt} the formation of 
clumps at the RD epoch from {\em adiabatic\/} spiky  perturbations. The 
difference to isothermal  perturbations is mainly in their evolution during 
the linear stage: While isothermal  perturbations are frozen in,
adiabatic fluctuations grow logarithmically. 

We include in this and in the accompanying paper~\cite{pII} 
a discussion of the detection prospects of
stable superheavy DM particles. Since the annihilation signal from the mean
distribution of these particles in the halo is far below observational limits,
we examine whether there are new effects which improve
the detection chances. One such effect follows from the early kinetic decoupling of 
superheavy DM particles from the thermal plasma. In this case 
the cutoff mass can be significantly smaller, as e.g.\ in the case of 
ultra-cold WIMPs~\cite{GelGon08}, and thus clumps of practically 
all masses are formed.
This opens the door for the formation of light superdense  DM clumps at the
RD stage. The only 
necessary condition is the existence of  spiky small-scale perturbations.

This article is organized as follows. We determine the initial properties 
of the DM clumps, first assuming
a standard power-law for the initial cosmological perturbations in
Sec.~\ref{params} and then a spiky perturbation spectrum in
Sec.~\ref{sec:spiky}. In Sec.~\ref{pbhs}, we derive constraints on the
superdense clump scenario considering primordial black hole production.
Then we study the evolution of the density profile of superdense clump
in Sec.~\ref{dprofevols}, commenting on the case of neutralinos
with masses close to the electroweak mass scale in Sec.~\ref{EWneutralino}.
We present finally our conclusions  in Sec.~\ref{sec:conc}.

%%%%%%%%%%%%%%%%%%%%%%%%%%%%%%%%%%%%%%%%%%%%%%%%%%%%%%%%%%%%%%%%%%%%%%%%%
\section{Clumps in the standard cosmological scenario}
\label{params}

We briefly remind in this section the formation of clumps and their 
properties
assuming a standard power-law spectrum of the initial cosmological 
perturbations. In contrast to the
usual approach, we allow here very small masses of the clumps 
being inspired by the smallness of 
$M_{\rm min}$ in the case of superheavy DM (SHDM), where  
$M_{\rm min}$ can be of order of  SHDM particle mass $m$. 

\begin{figure}%[t]
\begin{center}
\includegraphics[angle=0,width=0.5\textwidth]{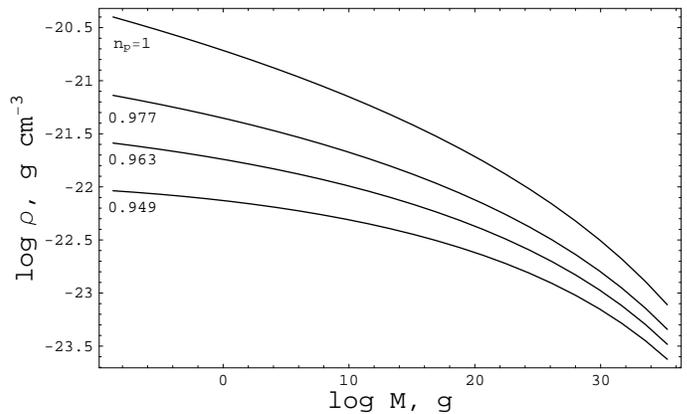}
\end{center}
\caption{The mean density $\rho$ of DM clumps as function of the clumps 
mass $M$ for different spectral indices $n_p$ of the primordial density 
perturbations.} \label{gr1}
\end{figure}
\begin{figure}[t]
\begin{center}
\includegraphics[angle=0,width=0.5\textwidth]{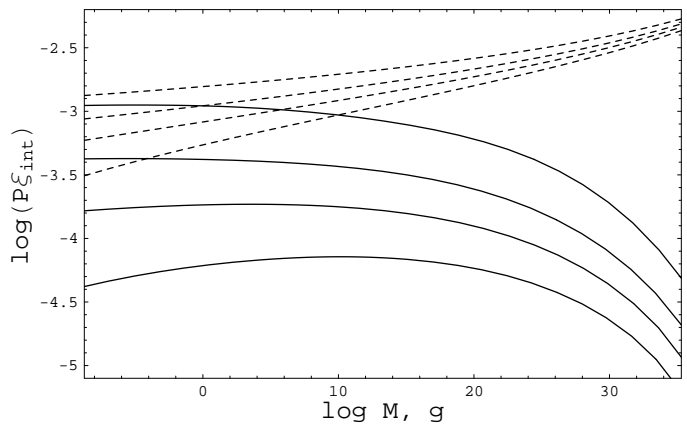}
\end{center}
\caption{The fraction of DM in the form of survived clumps per 
unit logarithmic
mass interval $\delta M\sim M$ as function of of clump mass $M$ for
$n_p=0.949$ (bottom), $0.963$, $0.977$ and $1$ (top): The initial fractions
are shown by dashed lines, the present fractions by
solid lines.} \label{gr2}
\end{figure}

Small clumps form at the MD epoch for $z\gg1$, i.~e.\ at a time when the
effect of the cosmological constant can be still neglected. In the
spherical model of the Press-Schechter theory~\cite{ps74,cole}, the
formation of an object occurs at the time $t_f$ when the density contrast
$\delta(M,t_f)$ reaches $\delta_c=3(12\pi)^{2/3}/20\simeq1.686$. The
mean density $\bar\rho_{\text{int}}$
and the radius $R$ of the collapsing clumps are
\begin{equation}
\bar\rho_{\text{int}}=\kappa\bar\rho(z_f)= \kappa\rho_{\text{\rm
eq}}\left(\frac{1+z_{f}}{1+z_{\text{\rm eq}}}\right)^3=
\kappa\rho_{\text{\rm eq}}\frac{\nu^3\sigma_{\text{\rm
eq}}^3(M)}{\delta_c^3}, \label{rhocl}
\end{equation}
and
\begin{equation}
  R=\left(\frac{3M}{4\pi\bar\rho_{\text{int}}}\right)^{1/3},
\label{rad1}
\end{equation}
where $\kappa = 18\pi^2 \simeq 178$~\cite{cole}, $\sigma_{\rm
eq}(M)$ is the variance and $\nu=\delta_{\rm eq}/ \sigma_{\rm
eq}(M)$ is the peak height of the density fluctuations at the time
$t_{\rm eq}$ of matter and radiation equality, while $\rho_{\text{\rm
eq}}$ is the density at $t_{\rm eq}$.

According to Ref.~\cite{BDE08}, surviving clumps are characterized
by $\nu\simeq 1-3$ and we set $\nu=2$ in all following
calculations. Having fixed $\nu$, the dependencies $R(M)$ and
$\bar\rho(M)$ are unambiguous and the  mean density $\bar\rho$ of
small-scale DM clumps as function of the clumps mass $M$ is shown 
in Fig.~\ref{gr1}. 

The mass function of clumps, i.e.\ the fraction of DM in the form of
clumps with mass $M$, is given by~\cite{BDE08}
\begin{equation}
 \xi_{\rm int}\frac{dM}{M}\simeq0.02(n+3)\,\frac{dM}{M} \,,
 \label{xitot}
\end{equation}
where the effective exponent $n$ in Eq.~(\ref{xitot}) is
found  as $n=-3(1 - 2\partial\log\sigma_{\rm eq}(M)/\partial\log
M)$ and depends very weakly on $M$. The simplest inflation models
give $P(k)\propto k^{n_p}$ with $n_p \approx 1$. The 7-year WMAP
data, $n_p=0.963\pm 0.014$, favor clearly $n_s<1$~\cite{WMAP7}.
Clumps can form nevertheless, because of the presence of
additional logarithms in the transfer function. The small-scale
spectrum at the epoch of matter-radiation equality can be
expressed as~\cite{bde03} 
\ba
 \sigma_{\text{\rm eq}}(M) &\simeq & 8.2\times 10^{3.7(n_p-1)-3}
 \left(\frac{M}{M_{\odot}}\right)^{\frac{1-n_p}{6}}
\nonumber\\ & \times &
 \left[1-0.06\log\left(\frac{M}{M_{\odot}}\right)\right]^{\frac{3}{2}} \,.
\label{A5}
\ea

The mass function (\ref{xitot}) with the spectrum (\ref{A5}) is
shown in Fig.~\ref{gr2} by dashed lines. Its $1/M$ shape is in
good agreement with the corresponding numerical simulations of
Ref.~\cite{DieMooSta05}, only its normalization is a few times
smaller than the one found there. For an extrapolation by many 
orders of magnitudes this must be considered as remarkable agreement.

Note also that using the power-law
spectrum  that is normalized to the temperature fluctuations of
the CMB, i.e.\ at cosmological scales, for sub-galactic scales or
even DM clumps with mass $M\sim 1 {\rm g}$ implies an 
extrapolation by
$\sim 48$ orders of magnitudes. This extrapolation can be
justified only within the simplest models for inflation.

Integrating  the mass
function (\ref{xitot}) from $M_{\rm min}$ to $M\sim10^2M_{\odot}$,
we obtain the {\em initial\/} (i.e.\ before possible destruction 
in the Galaxy)  fraction of DM in the form of clumps. 
In contrast to the standard case of EW scale neutralinos, where 
$M_{\rm min} \sim (10^{-6} - 10^{-8})M_{\odot}$ \cite{bde06}, 
in superdense 
clumps the DM particle can have much larger mass and thus 
$M_{\rm min}$ can be much smaller. As a result the fraction of 
surviving clumps increases. In particular for superheavy 
neutralinos, $M_{\rm min}$ can be comparable to the particle mass $m$
and the fraction of surviving clumps is calculated as 
as  0.15, 0.18, 0.21 and 0.26 for $n_p=0.949$,
$0.963$, $0.977$ and $1$, respectively. 

\begin{figure}[t]
\begin{center}
\includegraphics[angle=0,width=0.5\textwidth]{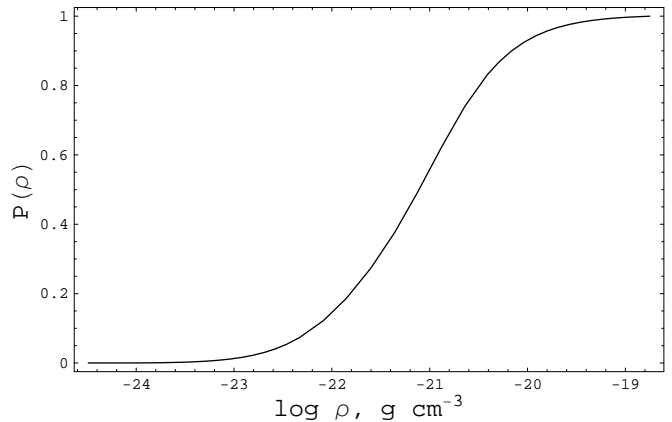}
\end{center}
\caption{The survival probability $P(\rho)$ as function of the mean 
internal clump density $\rho$ at the distance
$8.5$~kpc from Galactic center.} \label{gr3}
\end{figure}

Clumps inside galaxies lose mass and can be destroyed in
tidal interactions with stars. The collective gravitational field
of the Galactic disk is the most important factor for the clump
destruction. A method to study the destruction process of clumps
was presented in \cite{bde06} (for a more detailed approach with
gradual mass loss see \cite{BDE08}), where only clumps with
$M>10^{-6}M_{\odot}$ were considered. Here we calculate the
survival probability for the wider mass interval
$m<M\le10^2M_{\odot}$, using the same formalism as in
\cite{bde06}. The result for the survival probability $P(\rho)$ at
the position of the Sun, $r=8.5$~kpc from the Galactic center is
presented in Fig.~\ref{gr3}. Note that the survival
probability $P(\rho)$ means the fraction of surviving clumps 
near the Sun but most of these clumps have elongated orbits and spend
the largest part of their orbital period far from the Sun at the outer
parts of the Galactic halo.

The resulting mass function $P\xi_{\rm int}$ that accounts for the
effect of tidal destruction by stars is shown in Fig.~\ref{gr2}
by solid lines. Integrating $\int P\xi_{\rm int}dM/M$ again from
$M\sim m$ to $M\sim10^2M_{\odot}$ we obtain the actual fractions
of DM in the form of clumps as 0.006, 0.015, 0.033 and 0.085 for
$n_p=0.949$, $0.963$, $0.977$ and $1$, respectively.

\begin{figure}%[t]
\begin{center}
\includegraphics[angle=0,width=0.5\textwidth]{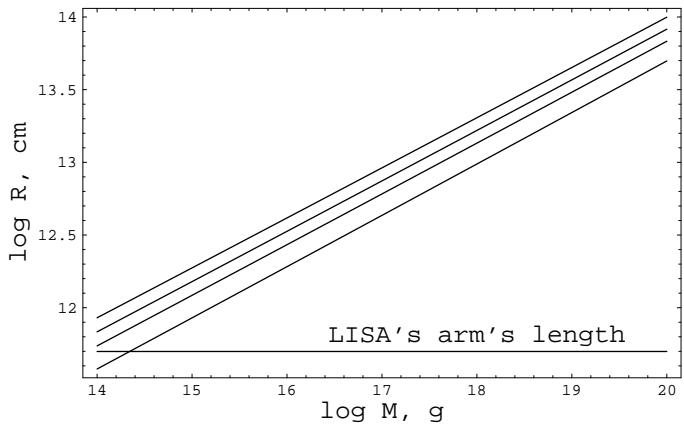}
\end{center}
\caption{The mean virial radius $R$ of DM clumps (\ref{rad1}) as function of the clump 
mass $M$ for several values of the spectral index of primordial density 
perturbations (from top to bottom): $n_p=0.949$,
$0.963$, $0.977$ and $1$. The horizontal line shows  LISA's
arm's length $L\simeq5\cdot10^{11}$~cm.} \label{gr4}
\end{figure}

Clumps formed from the standard power-law spectrum considered above 
have a rather small density. For many DM particle candidates,
including SHDM particles, such clumps are unobservable via their
annihilation signal and these clumps can be detected only gravitationally. 
It has been already suggested that interferometric detectors for 
gravitational waves like LISA have the capability to detect the tiny variation
of the gravitational field, when a compact object crosses the detector.
Small SHDM clumps should be included in the list of objects to be
searched for by LISA, such as  primordial black 
holes~\cite{LISAPBH}, asteroids~\cite{LISAAsteroid} or 
compact DM objects of unknown nature~\cite{LISADM}. 
The observable signal is caused by the gravitational tidal force which 
changes the interferometer arm length and produces correspondingly 
a phase shift. LISA will have the capability to search for compact
objects in the mass interval $10^{16}$~g$\le M\le10^{20}$~g
according to Ref.~\cite{LISAPBH} and $10^{14}$~g$\le M\le10^{20}$~g
according to Ref.~\cite{LISADM}. The signal will be in the form of
single pulses with its characteristic frequency at the lower end of
the expected LISA sensitivity curve and a rate $\sim$ a few per decade,
if the objects constitute the major part of DM. The clumps under
consideration in this Section present only 1-10\% of all DM, and
correspondingly, the detection rate will be 1-2 order of magnitudes
lower. In addition, the radii of the clumps generally exceed
LISA's arm length $L\simeq5\cdot10^{11}$~cm (see the Fig.~\ref{gr4}) and 
the tidal
forces will be smaller due to the extension of these objects.
Therefore, the detection of the SHDM clumps by LISA seems unlikely. The next
generation of gravitational wave interferometers offers
more promising perspectives for detection (for details see
Ref.~\cite{LISAPBH}). 

%%%%%%%%%%%%%%%%%%%%%%%%%%%%%%%%%%%%%%%%%%%%%%%%%%%%%%%%%%%%%%%%%%%%%%%%%%
\section{Non-standard perturbations and superdense clumps}
\label{sec:spiky}

%%%%%%%%%%%%%%%%%%%%%%%%%%%%%%%%%%%%%%%%%%%%%%%%%%%%%%%%%%%%%%%%%%%%%%%%%%
\subsection{Spiky density perturbation spectrum}
\label{spikys}

The variance of the normalized power-law spectrum at the horizon scale during
the RD stage was expressed for the standard inflationary scenario in
Ref.~\cite{GreLid97} as
\begin{equation}
\sigma_H(M)\simeq9.5\times10^{-5}\left(\frac{M}{10^{56}
\mbox{~g}}\right)^{\frac{1-n_p}{4}}.
\end{equation}
We see that in view of current observations ($n_p<1$) the variance
$\sigma_H(M)$ is too small for the formation of clumps at the RD stage.
Such clumps can be produced effectively only from non-standard spectra
containing e.g.\ spikes.

A sharp peak emerges in the fluctuation spectrum e.g., if an
inflationary potential $V(\phi)$ has a flat segment
\cite{star,ivan94}. The mean density perturbation on the horizon
scale is $\delta_{\mathrm{\rm H}}\sim
M_{\mathrm{Pl}}^{-3}V^{3/2}/V'$.
Hence, if the derivative $V'=dV(\phi)/d\phi\to0$ for some value
of the scalar field $\phi$, then a peak emerges in the
perturbation spectrum on the corresponding scale.  A similar
effect can arise in inflationary models with several scalar fields
\cite{yok95,gars96}.  In both types of models, the spectrum
outside the peak can have an ordinary shape. In particular, it can
be a Harrison--Zel'dovich spectrum, and can give rise to galaxies,
clusters and superclusters according to the standard scenario.

Another possibility to generate a spiky density perturbation spectrum
are cosmological phase transitions, for example
the QCD phase transition \cite{ssw}.
If somewhere  a high peak arises in the perturbation spectrum,
then the corresponding clumps would be the densest DM objects in
the universe. Theoretical models for nonstandard spectra were
discussed also in \cite{cline03}. A peak in $P(k)$ was proposed also in
\cite{tka99}. The authors of Ref.~\cite{DemDor03} found evidence for excess
power at small scales $\sim10h^{-1}$~kpc in comparison with a flat
primordial power spectrum. This result was obtained from the study of
Lyman-$\alpha$ absorbers and can be explained within complex
inflation models with the generation of extra power at small scales.
Such models can lead to the effective production of very dense clumps.

We will refer to all these models collectively as spiky models or 
spiky mass-spectrum models.

Dark matter clumps are formed in a wide range of masses, if the power
spectrum of primordial cosmological density perturbations has a
power-law form.  If on the contrary the spectrum has a peak on
some scale, then clumps are formed mostly in a narrow range of
masses, near the mass that corresponds to this peak.

%%%%%%%%%%%%%%%%%%%%%%%%%%%%%%%%%%%%%%%%%%%%%%%%%%%%%%%%%%%%%%%%%%%%%%%%%%
\subsection{Formation of superdense DM clumps at the RD epoch}
\label{forms}

A useful approximation for the nonlinear evolution of perturbations in
the radiation dominated epoch is the spherical collapse model~\cite{ind,kt}.
In this model, the evolution of perturbations after the horizon crossing
is described by
\begin{equation}
 y(y+1)\frac{d^2b}{dy^2} + \left[1 + \frac{3}{2}y\right]
 \frac{db}{dy} + \frac{1}{2} \left[ \frac{1+\Phi}{b^2}-b \right]
 = 0 \,,
 \label{bigeq}
\end{equation}
where $y=a(\eta)/a_{\rm eq}$, $\eta=dt/da$ is the conformal time,
$a_{\rm eq}$ is the scale factor at $\eta_{\rm eq}$,
and $\Phi = \delta\rho_{\rm DM}/\rho_{\rm DM}$ is the relative
overdensity of DM. The radius of the perturbed region is parametrized as
\begin{equation}
 r=a(\eta)b(\eta)\xi \,. \label{abxi}
\end{equation}
Here, $\xi$ is the comoving coordinate of the spherical layer considered
and the value $b(\eta)$ takes into account the slow-down of the
cosmological expansion in the perturbed density region.
Equation~(\ref{bigeq}) is applicable for the evolution of both
entropy and adiabatic perturbations, but has to be used with different
initial conditions.

The formation of clumps from entropy perturbations was considered in
\cite{kt}.  In this particular case, the initial data have the form
$\Phi=\delta\rho_{\rm DM}/\rho_{\rm DM}$ and $db/dt=0$.
The object formed has the density \cite{kt}
\begin{equation}
 \rho\simeq140\Phi^3(\Phi+1)\rho_{\rm eq}.
 \label{rho140}
\end{equation}
For instance, $\Phi\simeq 1\div10^4$ in the case of axions as DM,
and axionic miniclusters have masses in the range
$\sim(10^{-13}\div0.1)M_{\odot}$.  The observational
signatures of the presence of these axionic miniclusters in the
Galactic halo were considered in \cite{kt,fl}.

The corresponding method for the nonlinear evolution of adiabatic
perturbations during the radiation dominated epoch is described in
\cite{dokero2002}. For adiabatic perturbations $\Phi=0$, the
initial velocity $db/dt$ is nonzero and is defined using linear
perturbation theory. The transformation from the Euler
description for the growth of density perturbations $\delta$ to the
Lagrange description (\ref{abxi}) is provided by the relation
$b=(1+\delta)^{-1/3}$ \cite{ind}. The evolution of perturbations with
$\delta\ll1$ on scales less than the horizon is defined by the known
analytic solution \cite{ssw} (see also \cite{dokero2002})
\begin{equation}
 \delta=\frac{3A_{\rm in}}{2}\left[
 \ln\left(\frac{x}{\sqrt{3}}\right)
+\gamma_E-\frac{1}{2} \right] \,.
 \label{dgame}
\end{equation}
In this solution the numerical constant equals
$\gamma_E-1/2\approx0.077$, $A_{\rm in}=\delta_{\rm H}/\phi$,
$\phi\simeq0.817$, $\delta_{\rm H}$ is the radiation density
perturbation on the horizon scale and the variable $x$ is related to 
the comoving wave-vector $k$ of the  perturbation by $x=k\eta$. The
connection between $x$ and $y$ is defined by the relation
\cite{dokero2002}
\begin{equation}
 x=\frac{\pi}{2^{2/3}}\left(\frac{3}{2\pi}\right)^{1/6}
 \frac{cy}{M^{1/3}G^{1/2}\rho_{\rm eq}^{1/6}} \,.
\end{equation}
It is suitable to connect the analytic solution of  the linear theory
(\ref{dgame}) with the numerical solution of the nonlinear
Eq.~(\ref{bigeq}) at the time corresponding to the ``transition''
value of perturbations with  $\delta=0.2$ (see \cite{dokero2002}).  At
this moment we define the initial velocity of the forming DM clump as
\begin{equation}
 \frac{db}{dy}=-\frac{\delta_{\rm H}b^4}{2y\phi} \,.
\end{equation}

\begin{figure}[t]
\begin{center}
\includegraphics[angle=0,width=0.45\textwidth]{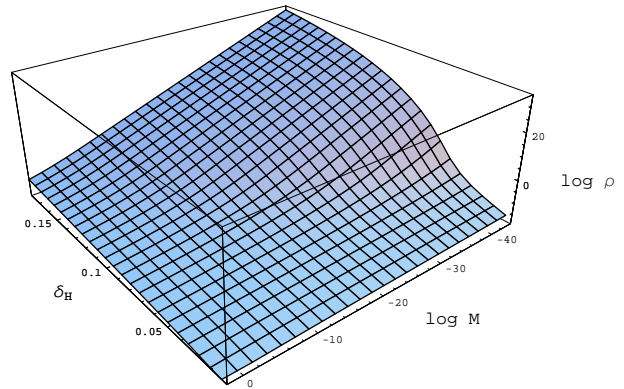}
\end{center}
\caption{The  mean density $\rho$ (in g~cm$^{-3}$) of DM clumps as function 
of the perturbation $\delta_{\rm H}$ in the radiation density on the horizon 
scale and the clump mass $M$ (in $M_{\odot}$).} \label{gr5}
\end{figure}

The cosmological expansion of the forming DM clump stops when
$dr/dt=0$ or according to Eq.~(\ref{abxi}) when
$db/dy=-b/y$.  The corresponding density and radius of the clump are
\begin{equation}
 \rho_{\rm max}=\rho_{\rm eq}y_{\rm max}^{-3}b_{\rm max}^{-3}, \quad
 R_{\rm max}=\left(\frac{3M}{4\pi\rho_{\rm max}}\right)^{1/3} \,,
 \label{rmax}
\end{equation}
where $b_{\rm max}$ and $y_{\rm max}$ are respectively the values
of $b$ and $y$ at the same moment.  After decoupling from the
cosmological expansion, the object virializes and contracts
by a factor two. In Ref.~\cite{dokero2002} this model was used to 
describe a noncompact 
DM object with single mass $\sim0.1M_\odot$, presumably observable
through microlensing. 
Now we consider the whole possible range of masses and densities of DM clumps.  
Calculating numerically the solution of Eq.~(\ref{bigeq}) within the above
formalism, we find the density of the clump $\rho=\rho(M,\delta_{\rm H})$ as function
of its mass $M$ and the radiation perturbation value on
the horizon scale $\delta_{\rm H}$ as shown in Fig.~\ref{gr5}.

\begin{figure}[t]
\begin{center}
\includegraphics[angle=0,width=0.45\textwidth]{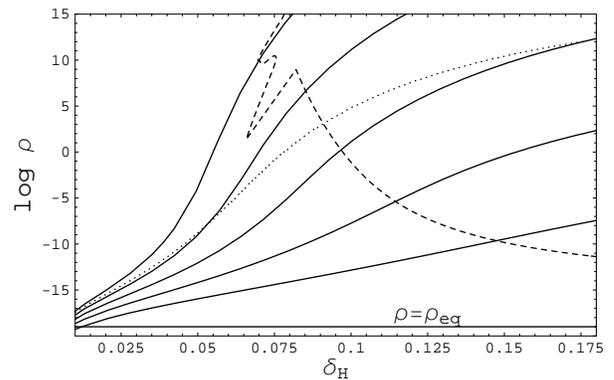}
\end{center}
\caption{The mean density $\rho$ (in g~cm$^{-3}$) of DM clumps as function 
of the perturbation $\delta_{\rm H}$ in the radiation density on the horizon 
scale; solid lines from top to bottom are for clump masses $M=10^{-10}$, 
$1$, $10^{10}$, $10^{20}$, $10^{30}$~g. The dashed line is the bound on
the clump density from primordial black holes overproduction with threshold 
$\delta_c=0.7$. The time of two-body gravitational relaxation inside the 
clump core is less than the age of the Universe for clumps above
the dotted lines, if the DM particle mass is $m\geq 10^{11}$~GeV.
\label{gr6}}
\end{figure}

Some characteristic values of the clump density $\rho$ are displayed
in Fig.~\ref{gr6} for several values of the clump mass $M$. 
One observes the convergence of curves to $\rho\sim\rho_{\rm
eq}\sim10^{-19}$~g~cm$^{-3}$ at small $\delta_{\rm H}$, i.e.\ for
clumps formed near matter-radiation equality. This corresponds to the
known analytical results that the evolution during the MD epoch does not
depend on the mass but only on the initial (at $t=t_{\rm eq}$) value
of the fluctuation.

Note that (in contrast to the case with standard power-law
spectrum of cosmological perturbations) superdense clumps from a
spike in the spectrum are not destroyed by tidal forces and their
mass function peaks near a definite mass. Therefore the
fraction of DM in the form of such clumps is $\xi\sim1/2$. Half
of the volume is in the form of overdensities (clumps), and the 
remaining space is filled by voids. Because of the compactness of 
superdense clumps, these clumps can satisfy the condition 
$R<L\simeq5\cdot10^{11}$~cm for the mass interval 
$10^{14}$~g$\le M\le10^{20}$~g and are thus   
observable by the LISA detector.

%%%%%%%%%%%%%%%%%%%%%%%%%%%%%%%%%%%%%%%%%%%%%%%%%%%%%%%%%%%%%%%%%%%%%%%%%%
\section{Clumps and primordial black holes}
\label{pbhs}

The formation of DM clumps leads to several restrictions on the fluctuation 
spectrum. For instance, high-energy particles from DM annihilations
in clumps during the epoch of nucleosynthesis and after it (the cascade 
nucleosynthesis) might distort the prediction of standard
nucleosynthesis. 

Another important restriction on the spectrum of the adiabatic
perturbations comes from upper limits on the mass and density
of primordial black holes (PBHs) \cite{zeld67,carr75},
because the value of DM density perturbations depends on the
radiation density perturbations and the formation of DM
clumps can be related to the formation of PBH from the same
perturbation spectrum \cite{dokero2002}. It should be noted that in
the case of entropy perturbations PBHs do not form.

Clumps and PBHs originate from fluctuations of the same
type but are formed at different times. The large
difference between the masses of DM clumps and of PBH
arises because of the large difference in energy density enclosed in a fixed
comoving volume as function of time: The energy density of
radiation at the RD epoch  far exceeds the 
mass in CDM at the matter domination epoch in the same comoving volume.

The formation of PBHs takes place on the tail of the distribution of
Gaussian fluctuations, whereas the main part of clumps is produced
from r.m.s. fluctuations.  Therefore only a small part of the
fluctuations which result in the formation of clumps may produce PBHs
at the RD epoch.  In other words, because of the large threshold of 
PBH formation, the major part of fluctuations does
not collapse into PBHs and evolves continuously up to the end of
the RD epoch.  During the RD epoch the
mass of radiation in the comoving volume varies as $M_r(t) = M
a(t_{\rm eq})/a(t)$, where the scale-factor of the Universe
$a\propto t^{1/2}$ and $M$ is the comoving mass at the moment of
transition to MD.  The mass $M$ equals
approximately to the mass of non-relativistic matter inside the
fluctuation, i.e.\ the mass of a clump which may be formed from this
fluctuation.  On the other hand at horizon crossing
$M_r(t)\sim4\pi(2ct)^3\rho(t)/3$, where $\rho(t)=3/32\pi Gt^2$.
>From these relations for $M_r(t)$ we estimate the mass $M_H$ and
the time $t_H$ of PBH formation as function of the clump mass $M$ as
\begin{equation}
M_H\sim cM^{2/3}G^{-1/3}t_{\rm eq}^{1/3},~~~~~t_H\sim GM_H/c^3 \,.
\label{mhmx1}
\end{equation}
From the Friedmann equations, the formula for $M_H$ was derived
exactly in Ref.~\cite{dokero2002} and is given by
\begin{equation}
 M_{\mathrm{\rm H}} =
 \frac{1}{2^{2/3}}
\left(\frac{3}{2\pi}\right)^{1/6}
 \frac{M^{2/3}c}{G^{1/2}\rho_{\mathrm{\rm eq}}^{1/6}}
 = 2\times10^5
\left(\frac{M}{0.1M_{\odot}}\right)^{2/3} M_{\odot} \,, \label{mhmx}
\end{equation}
while the dependence $t_H(M)$ is 
\begin{equation}
 t_{\mathrm{\rm H}}
  = 3.7\left(\frac{M}{M_{\odot}}\right)^{2/3}
\mbox{~s}. \label{thmx}
\end{equation}

The fraction of the mass in radiation that is transformed into PBHs at
the time $t_{\mathrm{H}}$ is then \cite{carr75}
\begin{equation}
\beta= \int\limits_{\delta_{\mathrm{c}}}^{1} \frac{\displaystyle
d\delta_{\mathrm{H}}}{\displaystyle\sqrt{2\pi}\Delta_
{\mathrm{H}}}
\exp(-\frac{\displaystyle\delta_{\mathrm{H}}^2}{\displaystyle2\Delta_{\mathrm
{H}}^2}) \simeq
\frac{\Delta_{\mathrm{H}}}{\delta_{\mathrm{c}}\sqrt{2\pi}}\exp(-\frac{\delta_
{\mathrm{c}}^2}{2\Delta_{\mathrm{H}}^2}), \label{bet2}
\end{equation}
where $\delta_c$ is the threshold value of the density
perturbations $\delta_H$ which result in PBHs formation.  The
current PBHs density parameter is $\Omega_{\mathrm{BH}}\simeq\beta
a(t_{\mathrm{\rm eq}})/a(t_{\mathrm{H}})$.

For a large enough value of the r.m.s.\  perturbation $\Delta_{\rm H}
\equiv \langle\delta_{\rm H}^2\rangle^{1/2}$, an extremely large
number of PBHs can be formed \cite{carr75}.  This provides a
limitation on $\Delta_{\rm H}$.

The number density of PBHs depends strictly on the
threshold value $\delta_c$.  In early works, e.~g. \cite{carr75,nad78,nov79}, 
the value of $\delta_c=1/3$ was
obtained.  In recent years the phenomenon of critical
gravitational collapse was discovered in numerical
simulations, for which $\delta_c\simeq0.7$ \cite{chop,niem98}.
Some limits on the number density of PBHs in different mass ranges
were obtained in \cite{carr75,nov79}. These restrictions on the
value of $\Delta_{\rm H}$ for PBHs are shown in Fig.~2 for the
case $\delta_c=0.7$. The relation (\ref{mhmx}) was used in our
calculations. The local minimum on the curve corresponds to the
restrictions on the Hawking evaporating PBHs with masses $M_{\rm BH}
\simeq 10^{15}$~g. For PBHs with a larger mass the only
restrictions comes from the condition that their cosmological
density parameter $\Omega_{\rm PBH}\le1$.

We recall that PBHs are formed on the tail of the Gaussian perturbation
distribution, $\delta_{\rm H} \ge \delta_c\gg\Delta_{\rm H}$.  On
the contrary the overwhelming number of DM clumps are formed from
the r.m.s perturbations.  For this reason in Fig.~2 and in the
calculations for DM clumps we put $\delta_{\rm H}\simeq\Delta_{\rm
H}$.

%%%%%%%%%%%%%%%%%%%%%%%%%%%%%%%%%%%%%%%%%%%%%%%%%%%%%%%%%%%%%%%%%%%
\section{Superdense clumps from ordinary neutralinos}
\label{EWneutralino}

We consider in this section the standard case of thermally produced 
neutralinos with mass close to the electroweak mass scale. We will
show that the diffuse gamma flux produced by such neutralinos constituting
superdense clumps exceeds the observed flux, and thus superdense clumps
should consist of DM particles non-thermally produced. 

In order to make our estimate most transparent, we consider first the
integral photon flux produced by DM annihilations. This flux is easy to 
estimate using the annihilation cross section $\langle\sigma v\rangle$ 
for the process $\chi + \chi \to \pi^0 +$all and the mean density 
$\bar{\rho}_{\rm int}$ of neutralinos in a clump.

We calculate first the rate $\dot{N}_{\gamma}$ of gamma-rays with energies 
higher than 70~MeV produced by a single clump, assuming a $r^{-1.8}$ density 
profile with core at $r \leq R_c$ for a clump with total mass $M$ and 
radius $R$,
\be 
\dot{N}_{\gamma} = 1.6 \eta_{\pi^0} \frac{\langle\sigma v\rangle}{m_\chi^2}
\bar{\rho}_{\rm int} M \left (\frac{R}{R_c} \right )^{0.6} \,,
\label{N-gamma}
\ee
where $\eta_{\pi^0}$ is neutral pion multiplicity, 
$R_c=x_cR$ is the core radius, and $m_\chi$ is the neutralino mass.   

The total diffuse flux produced in a  galactic DM halo can be 
calculated as 
\be
J_{\gamma}^{\rm tot} = \frac{1}{4\pi} \bar{n}_{\rm cl} R_h \dot{N}_{\gamma}
\label{diff-flux-gen}
\ee
with $R_h$ as the radius  of the DM halo and $\bar{n}_{\rm cl}$ as the clump
mean space density, which is given by the fraction $\xi$ of DM in the form 
of clumps and the mass $M_h$ of the galactic DM  halo as  
\be 
\bar{n}_{\rm cl}= \frac{3}{4\pi} \frac{\xi M_h}{M} \frac{1}{R_h^3} \,.
\label{n_s}
\ee
Using Eqs.~(\ref{N-gamma}-\ref{n_s}) we can express the diffuse flux 
$J_{\gamma}^{\rm tot}$ in terms of the mean density of neutralinos in
clumps 
$\bar{\rho}_{\rm int}$, the main characteristic of superdense clumps:
\be
J_{\gamma}^{\rm tot}= f_{\rm NFW}\frac{0.4}{\pi}\frac{\eta_{\pi^0} \xi R_h}{x_c^{0.6}} 
\frac{\bar{\rho}_{\rm int}\bar{\rho}_{\rm halo}^{DM}} {m_\chi^2} \langle\sigma v\rangle,
\label{tot-flux}
\ee
where $\bar{\rho}_{\rm halo}^{DM}$ is the mean density of DM halo. To take into account 
the NFW density profile one must multiply the homogeneous halo result by the additional factor $f_{\rm NFW}=293$.  
The obtained flux is given as convenient expression where most of parameters
are observationally known and the main characteristic of superheavy
clumps is $\bar{\rho}_{\rm int}$.  For the  parameters in 
Eq.~(\ref{tot-flux}), we use $\eta_{\pi^0}=10$ appropriate for gauge boson decays, $\xi = 1/2$ (for
spiky scenario), $R_h \approx 200$~kpc
and $\bar{\rho}_{\rm halo}^{DM}=1.1\times10^{-3}$~GeV/cm$^3$ obtained as $3M_h/4\pi R_h^3$. 
We assume $x_c \approx 0.1$. 

Typically superdense clumps have very large densities (see Fig. \ref{gr6}) 
and for ordinary neutralinos the resulting gamma-ray flux exceeds the 
observations. 
First we analyze the problem, whether ordinary neutralinos are compatible
with a spiky scenario of clump production. With this aim we choose 
in Eq.~(\ref{tot-flux}) parameters which minimize the flux.  
For the mean density of neutralinos in a clump, $\bar{\rho}_{\rm int}$, we take 
the minimum value, assuming neutralino produced at the beginning of the MD
epoch. In this case  
$\bar{\rho}_{\rm int}= 178 \rho_{eq} = 8.3\times 10^6$~GeV/cm$^3$ (see Eq.~(\ref{rhocl}) in the limit $z_f\to z_{\rm eq}$). 
We parametrize the annihilation cross-section $\langle\sigma v\rangle$ 
by the characteristic value $1\times 10^{-26}$~cm$^3$/s as 
\be 
\langle\sigma v\rangle_{26} = \langle\sigma v\rangle/(10^{-26}~{\rm cm}^3 {\rm s}^{-1}) .
\ee
With this parameters the minimum gamma-ray flux is 
\be 
J_{\gamma}^{\rm tot} = 4.3 \langle\sigma v\rangle_{26} m_{100}^{-2} ~~~ {\rm cm}^{-2} 
{\rm s}^{-1} {\rm sr}^{-1}, 
\label{flux-num1}
\ee
where $m_{100}$ is the neutralino mass $m_\chi$ in units of 100 GeV.
 
The integral flux (\ref{flux-num1}) is about 5 orders of magnitudes larger
than the observed flux. Does it help to increase the neutralino mass or
to consider a smaller annihilation cross section? To answer the first
part of this question, we consider now the {\em differential\/} isotropic 
diffuse photon flux observed by Fermi-LAT at $|b|>60$~degrees \cite{TeV},
\be
 J_{\rm obs}(E) = 6\times 10^{-7} \left(\frac{E}{\rm GeV}\right)^{-2.45} {\rm GeV^{-1}\;cm^{-2}\;s^{-1}\;sr^{-1}} \,.
\ee
The differential photon flux produced by annihilations can be obtained from
Eq.~(\ref{tot-flux}) replacing $2\eta_{\pi^0}$ by $dN/(2m_\chi dx)$, where
$dN/dx$ is the number of photons with energy $E=xm_\chi/2$ produced per
annihilation. Since moreover $dN/dx$ increases at small $x$ for increasing
$m_\chi$, the ratio $J_{\gamma}(E)/J_{\rm obs}(E)$ is practically constant.
The minimal allowed annihilation cross-section of neutralinos 
obtained in \cite{berez96}   for the case of strongly suppressed $s$-wave
annihilations is 
$\langle\sigma v\rangle= 1.7\times 10^{-30}m_{100}^{-2} $~cm$^3$/s.
With these parameters the minimum gamma-ray flux  is
still above the measured Fermi-LAT flux.
Finally, we remark that even postulating at tree-level only couplings e.g.\
to electrons would lead to an overproduction of photons via
Bremsstrahlung, c.f.\ e.g. Ref.~\cite{brems}.
In conclusion, ordinary neutralino as other thermally produced DM particles
are excluded as constituents of superdense clumps. A DM particle suitable
to compose superdense clumps must have a smaller annihilation
cross section than allowed for a thermal relic.

%%%%%%%%%%%%%%%%%%%%%%%%%%%%%%%%%%%%%%%%%%%%%%%%%%%%%%%%%%%%%%%%%%%%%%%%%%%%%%
\section{Density profile evolution}
\label{dprofevols}

In the case of a spiky spectrum, clumps are formed  not in the 
process of hierarchical clustering but due to the evolution of isolated 
density fluctuations. Such a scenario is similar to the analytic approach      
in Ref.~\cite{ufn1}. The ordinary gravitational contraction combined with 
the multi-stream instability produces the universal power-law
density profile with exponent $\beta=1.7 - 1.9$ \cite{ufn1}.
This power-law shape for $\rho(r)$ has been recently confirmed in the
numerical simulations \cite{MDSQ} for neutralino clump formation during 
the MD epoch. We assume here that the clumps
produced at the RD stage in the process of ordinary gravitational
contraction  have a  profile
$\rho(r)\propto r^{-1.8}$ for $R_c<r<R$ and $\rho(r)=\rho_c={\rm
const}$ for $r<R_c$, where $R_c$ is the unknown core radius
of the clump. This core may be produced due to  tidal 
forces \cite{bde03} in the clumps formed at the MD stage. In this case 
a large core is produced with  $x_c\equiv R_c/R\sim0.01 - 0.1$.
More precisely, the given value corresponds not to the radius of the 
constant-density core but to the break in the slope of the
density profile. Moreover, the above-mentioned calculations are  
valid for the MD dominated epoch, where the process of core formation can
be much different from that at the RD epoch. 

Another estimate for the core size has been obtained in Ref.~\cite{ufn1}, 
where $x_c$ is defined by the damping mode of the perturbations. 
The authors obtained  $x_c\sim\delta_{\rm eq}^3$, where $\delta_{\rm eq}$ is the value of
density fluctuation at the beginning of the MD stage. 
However, this estimate is also valid for the MD epoch. 

At the current level of knowledge, the relative radius $x_c\equiv R_c/R$ 
of the core produced by ordinary gravitational contraction 
must be considered as a free parameter. In the
most conservative case we use $x_c\sim0.1$. The central density
$\rho_c$ depends on the mean clump  density $\bar\rho=8\rho_{\rm
max}$ (see (\ref{rmax})) as $\rho_c=\bar\rho/(3x_c^2)$.

We shall briefly discuss the evolution of  superdense clumps formed
from superheavy particles. Quantitatively, it will be considered in
the accompanying paper II.  

The first stage of evolution, the ordinary gravitational contraction, 
proceeds like in the case of ordinary neutralinos and results in the production 
of a $\rho (r) \sim r^{-1.8}$ profile with a relatively large core, 
$x_c \sim 0.01 - 0.1$. Other processes can become important at 
the second stage: (i) two-body gravitational scattering and (ii) some 
limiting effect like Fermi degeneracy or the intensive annihilation of 
particles.  In the cores of superdense clumps with large
densities $n$ of particles the binary gravitational scattering of
constituent DM particles with large masses $m$ may become the dominant
process, which causes the ``gravithermal instability'' or 
``gravithermal catastrophe'', well known in theory of globular star clusters. 
Note that this effect takes place 
only for superheavy DM and only in the most dense parts of  superdense 
clumps. In Fig.~\ref{gr6}, this region is located above the dotted line.

How can it be that gravitational two-body scattering becomes the
dominant process? It occurs because gravitational scattering is
proportional to $m^2$, while EW scattering of these 
particles is proportional to $1/m^2$. The other two factors are the large 
density $n$ of particles in the core and the long-range character of
gravitational interactions. All this provides the fast gravitational
relaxation of the system. As a result of the gravithermal instability a
clump develops an isothermal density profile $\rho (r) \propto r^{-2}$ 
with a tiny core. This core can be produced by the pressure of a 
degenerate gas in the case of superheavy fermionic particles or by
the inverse flow due to the annihilation 
of particles in the clump center \cite{berez96,bergur}. In these 
cases the radius of the new core is determined by the elementary particle 
properties of dark matter.

%%%%%%%%%%%%%%%%%%%%%%%%%%%%%%%%%%%%%%%%%%%%%%%%%%%%%%%%%%%%%%%%%%%%%%%
\section{Conclusions}
\label{sec:conc}

Superdense clumps can be produced from isothermal perturbations \cite{kt} 
or from spikes in the spectrum of adiabatic perturbations 
\cite{star,dokero2002}. These objects are produced 
in the very early universe during the RD epoch. In principle, the perturbation 
spectrum may include both a scale-invariant power-law component and spikes. 
Being produced very early during the radiation dominated epoch,
superdense clumps evolve as isolated objects. They do not belong 
to hierarchical structures for a long time after production, and 
therefore they are not destroyed by tidal interactions during the 
formation of large-scale structures. 

In the case of EW scale mass particles, e.g.\ ordinary neutralinos,
the density profile has a $r^{-1.8}$ shape with a relatively large core
characterized by $R_c/R \sim 0.01 - 0.1$, produced by tidal forces. 
Ordinary neutralinos are excluded as the constituents of superdense clumps, 
because they overproduce 
the diffuse gamma-ray spectrum above 100 MeV. The constituent DM
particles in superdense clumps must be either very weakly annihilating or be 
superheavy, or both. The limit on the  superdense
clumps is imposed by primordial black holes which originated from 
the same perturbation spectrum. The allowed intrinsic densities of 
superdense clumps are shown in Fig.~\ref{gr6}. The formation of superdense 
clumps at the RD epoch was studied previously using somewhat different 
assumptions in Refs.~\cite{kt,dokero2002,RicGou09}.

The density profile in superdense clumps depends on the properties of the DM
particles. For very heavy constituent particles and large intrinsic densities 
of the clumps a  ``gravithermal catastrophe''
(instability) may develop in
superdense clumps. As a result the initial density profile turns into an
isothermal one, $\rho_{\rm int}(r) \propto 1/r^2$ , and the large initial core 
collapses into a tiny, very dense new core.  The steep density
profile and the smallness of the core lead to a strong DM annihilation signal. 
The radiation produced by DM annihilations restricts this scenario,  e.g.\ 
due to the cascade nucleosynthesis following  standard nucleosynthesis.
On the positive side, superdense clumps can lead to detectable gamma 
radiation even in the case of superheavy DM particles~\cite{pII}. 
Superdense clumps can be in principle  observed also by gravitational 
wave detectors.

\acknowledgments
VB is grateful to the A.~Salam International Centre for Theoretical Physics
for hospitality during the work on this paper and to A.~Smirnov for valuable
discussions.
This work was supported by the Russian Federal Agency for Science and 
Innovation under state
contract 02.740.11.5092 and by the grants of the Leading scientific school 
959.2008.2 and 438.2008.2.

%%%%%%%%%%%%%%%%%%%%%%%%%%%%%%%%%%%%%%%%%%%%%%%%%%%%%%%%%%%%%%%%%%%%%%%%%%%%%%

\end{document}